\documentclass[lettersize,journal]{IEEEtran}
\usepackage{amsmath,amsfonts}
\usepackage{algorithmic}
\usepackage{algorithm}
\usepackage{array}
\usepackage[caption=false,font=normalsize,labelfont=sf,textfont=sf]{subfig}
\usepackage{textcomp}
\usepackage{stfloats}
\usepackage{url}
\usepackage{verbatim}
\usepackage{graphicx}
\usepackage{cite}
\usepackage{placeins}
\usepackage{amssymb}

\usepackage{pgfplots}
\usepgfplotslibrary{groupplots}
\usepackage{tikz}
\usetikzlibrary{patterns}
\usetikzlibrary{external}
\usetikzlibrary{spy}
\usepackage{multirow}

\newcommand{\ey}{\hat{\vec e}_{y}}
\newcommand{\ez}{\hat{\vec e}_{z}}

\renewcommand{\vec}[1]{\boldsymbol{#1}}

\newcommand{\paren}[1]{\left(#1\right)}

\newcommand{\volInt}[3]{\paren{#1 \, , #2}_{#3}}

\renewcommand{\div}{\textrm{div}\, }

\newcommand{\curl}{\textrm{\textbf{curl}}\, }
\newcommand{\curlOnly}{\textrm{\textbf{curl}}}

\renewcommand{\b}{\vec b}

\newcommand{\n}{\vec n}
\newcommand{\h}{\vec h}

\newcommand{\e}{\vec e}

\renewcommand{\j}{\vec j}

\newcommand{\dt}{\partial_t}

\newcommand{\happ}{\vec h_{\text{app}}}
\newcommand{\hmax}{h_{\text{max}}} 

\newcommand{\Ra}{R_{\text{a}}}
\newcommand{\Rc}{R_{\text{c}}}

\newcommand{\rhos}{\rho_{\text{s}}}
\newcommand{\Ns}{N_{\text{s}}}

\renewcommand{\O}{\Omega}

\newcommand{\Oc}{\Omega_{\text{c}}}
\newcommand{\Occ}{\Omega_{\text{c}}^{\text{C}}}
\newcommand{\Om}{\Omega_{\text{m}}}

\newcommand{\Os}{\Omega_{\text{s}}}

\newcommand{\hpf}{$h$-$\phi$-formulation\ }

\newcommand{\hpfOnly}{$h$-$\phi$-formulation}

\newcommand{\hsp}{\mathcal{H}}

\newcommand{\hspz}{\mathcal{H}_{0}}

\usepackage{eso-pic}

\usepackage{upgreek}

\usepackage{color}
\usepackage{xcolor}
\definecolor{myred}{rgb}{0.7,0.15,0.15}
\definecolor{mymaincolor}{rgb}{0.24, 0.36, 0.64}
\definecolor{mysecondcolor}{rgb}{0.21, 0.64, 0.87}
\definecolor{myblue}{rgb}{.2,0.45,0.5} 
\definecolor{myorange}{rgb}{0.78,0.6,0.3}
\definecolor{mygreen}{rgb}{.2,0.38,0.16}
\definecolor{myalert}{rgb}{0.97,0.09,0.21}
\definecolor{myformulation}{rgb}{0.33, 0.29, 0.31}
\definecolor{myformulation_back}{rgb}{1, 0.97, 0.91}
\definecolor{hf}{rgb}{0.93, 0.57, 0.13} 
\definecolor{hf_2}{rgb}{1.0, 0.89, 0.77} 
\definecolor{hf_3}{rgb}{1.0, 0.22, 0.0} 
\definecolor{hf_4}{rgb}{1.0, 0.4, 0.1} 
\definecolor{burlywood}{rgb}{0.87, 0.72, 0.53}
\definecolor{burntorange}{rgb}{0.8, 0.33, 0.0}
\definecolor{burntsienna}{rgb}{0.91, 0.45, 0.32}
\definecolor{af}{rgb}{0.4, 0.53, 0.34}
\definecolor{af_2}{rgb}{0.74, 0.77, 0.47}
\definecolor{af_3}{rgb}{0.12, 0.3, 0.17}
\definecolor{af_4}{rgb}{0.03, 0.34, 0.25}
\definecolor{haf}{rgb}{0.6, 0.51, 0.48}
\definecolor{haf_2}{rgb}{1, 0.97, 0.91}
\definecolor{taf}{rgb}{0, 0.55, 0.5}
\definecolor{ajf}{rgb}{0.29, 0.59, 0.82}
\definecolor{hbf}{rgb}{0.87, 0.36, 0.51}
\definecolor{prussianblue}{rgb}{0.0, 0.19, 0.33}
\definecolor{regalia}{rgb}{0.32, 0.18, 0.5}
	
\definecolor{myred}{rgb}{0.7,0.15,0.15}
\definecolor{mygreen}{rgb}{0.13,0.55,0.13}
\definecolor{myblue}{rgb}{0.25,0.41,0.88}

\definecolor{vir_0}{rgb}{0.993248, 0.906157, 0.143936}
\definecolor{vir_1}{rgb}{0.565498, 0.84243 , 0.262877}
\definecolor{vir_2}{rgb}{0.20803 , 0.718701, 0.472873}
\definecolor{vir_3}{rgb}{0.128729, 0.563265, 0.551229}
\definecolor{vir_4}{rgb}{0.190631, 0.407061, 0.556089}
\definecolor{vir_5}{rgb}{0.267968, 0.223549, 0.512008}
\definecolor{vir_6}{rgb}{0.267004, 0.004874, 0.329415}

\definecolor{bw_6}{rgb}{0.05, 0.03, 0.53}
\definecolor{bw_5}{rgb}{0.42, 0.  , 0.66}
\definecolor{bw_4}{rgb}{0.69, 0.17, 0.56}
\definecolor{bw_3}{rgb}{0.88, 0.39, 0.38}
\definecolor{bw_2}{rgb}{0.99, 0.65, 0.21}
\definecolor{bw_1}{rgb}{0.94, 0.98, 0.13}

	\definecolor{darkred}{rgb}{0.55, 0.0, 0.0}
	\definecolor{darkblue}{rgb}{0.0, 0.0, 0.55}

\usepackage{lipsum}  

\begin{document}

\title{Simulation of Rutherford Cable AC Loss and Magnetization with the Coupled Axial and Transverse Currents Method}

\author{Julien Dular, Fredrik Magnus, Erik Schnaubelt, Arjan Verweij, Mariusz Wozniak
\thanks{Authors are with CERN, Geneva, Switzerland. Fredrik Magnus is also with Norwegian University of Science and Technology, Trondheim, Norway. \textit{Corresponding author: Julien Dular.} Email: julien.dular@cern.ch.}
\thanks{Manuscript received 16 September, 2024.}}

{}


\maketitle

\begin{abstract}
The coupled axial and transverse currents (CATI) method was recently introduced to model the AC loss and magnetization in twisted composite superconducting strands with low computational cost and high accuracy. This method involves two-dimensional finite element (FE) models coupled with circuit equations representing the periodicity of the strand. In this paper, we propose to adapt the CATI method to Rutherford cables, which are periodic structures made of transposed superconducting strands. We focus on reproducing the interstrand coupling currents flowing across contact resistances between the strands and we analyze the associated AC loss. We show that results of a reference three-dimensional FE model are accurately reproduced with a strongly reduced computational cost.
\end{abstract}


\AddToShipoutPicture*{
    \footnotesize\sffamily\raisebox{0.4cm}{\hspace{1.5cm}\fbox{
        \parbox{\textwidth}{
            This work has been submitted to a journal for possible publication. Copyright may be transferred without notice, after which this version may no longer be accessible.
            }
        }
    }
}

\section{Introduction}
\IEEEPARstart{C}{lassical} three-dimensional (3D) finite element (FE) models of the magnetic response of transposed superconducting conductors such as Rutherford cables under transient conditions suffer from high computational costs, leading to impractically long simulation times~\cite{dangelo2021quasi, riva2023h}. Accurately modelling the response of the superconductor in terms of magnetization and AC loss is, however, crucial for designing magnet cryogenic systems, for assessing temperature and stability margins, as well as for developing novel quench protection techniques~\cite{wilson1983superconducting, mulder2023external, ravaioli2023optimizing}. To lower the computational cost compared to 3D FE models while still providing a sufficient accuracy and accounting for different conductor properties, reduced order models are necessary.

Various reduced order models for transient effects in Rutherford cables have been proposed. For example, network models describe the cable with lumped circuit elements~\cite{morgan1970theoretical, ries1981coupling, verweij1997electrodynamics},	 continuum models use distributed circuit elements along the cable length to describe coupling currents~\cite{krempasky1998influence, bottura2018calculation}. These models, however, do not account for the screening induced by coupling currents on the magnetic field distribution.

The coupled axial and transverse currents (CATI) method is a reduced order method based on two-dimensional (2D) FE models coupled with circuit equations~\cite{dular2024coupled}. It was recently proposed for modelling the transient magnetic response of twisted multifilamentary superconducting strands under magnetic field and transport current excitations. By reducing the dimension of the problem from 3D to 2D, the CATI method provides a substantial reduction of the computational cost compared to classical 3D FE models. It allows for an accurate treatment of the properties of the conductor cross-sections and it accounts for field distortions due to induced currents.

The CATI method only requires periodicity of the conductor to define the circuit equations and hence is applicable to the geometry of Rutherford cables. In this work, we adapt the CATI method approach to these cables. In particular, we focus on the interstrand coupling currents, flowing in-between different strands through contact resistances. We show that the CATI method can be used to accurately compute the associated magnetization and AC loss, verifying the results with those obtained with a classical 3D FE model.

In this verification analysis, we consider a linear problem in which strands are modelled as ohmic conductors with very low resistivity. This allows for a detailed analysis of the interstrand coupling current dynamics. Accounting for the nonlinear resistance and inductance of superconducting strands by coupling this cable model with a model describing small-scale dynamics in strands such as interfilament coupling currents or filament hysteresis is an important aspect and will be considered in further works.

Besides its low computational cost and good accuracy, one advantage of the CATI method is that it can be implemented in any FE program allowing for field-circuit coupling. Here, it is implemented in open-source and free to use software. Gmsh~\cite{gmsh} and GetDP~\cite{getdp} are used within the Finite Element Quench Simulator (FiQuS)~\cite{vitrano2023open} developed at CERN as part of the STEAM framework~\cite{Bortot2017}.

\section{CATI Method for Rutherford Cables}

\begin{figure}[h!]
\begin{center}
\includegraphics[width=\linewidth]{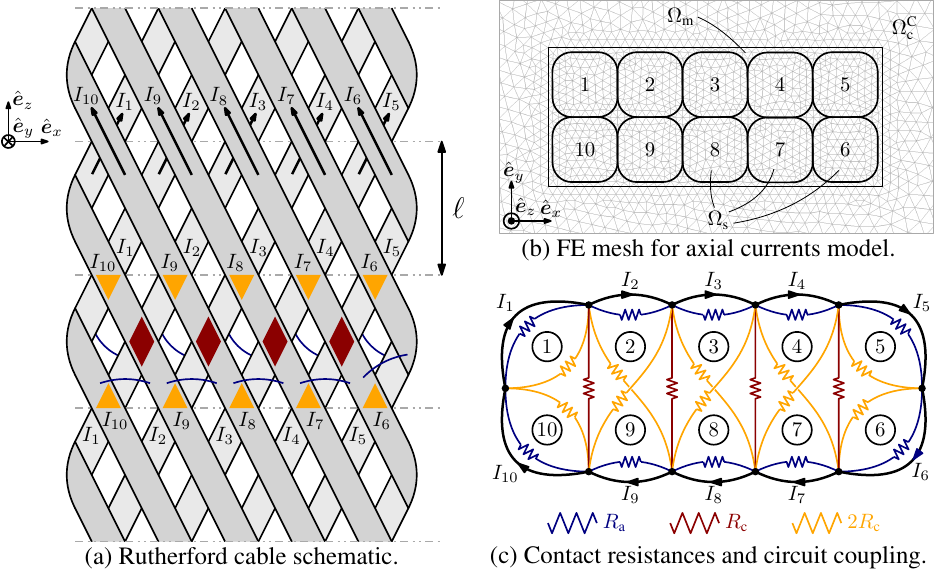}
\caption{Illustration of the CATI method on a Rutherford cable with $\Ns=10$ strands, with interpretation of the circuit coupling equations linking the axial and transverse currents models with adjacent ($\Ra$) and crossing ($\Rc$) contact resistances. The strands in (a) are represented without adjacent contacts for better visualization. See also Fig.~\ref{cable3D} for a 3D view. View (b) is a cross-section of the cable along one of the dash-dotted lines in (a), seen from above. In (b), the conducting domain $\Oc$ is the union of the strand domain $\Os$ and a coating domain $\Om$. Subfigure (c) shows the electrical circuit defining connections between the axial and transverse currents problems.}
\label{circuitModel}
\end{center}
\end{figure}

In this section, we introduce the CATI method equations for Rutherford cable modelling. We consider a cable as illustrated in Fig.~\ref{circuitModel} (see also Fig.~\ref{cable3D}), made of $\Ns$ composite superconducting strands transposed and compressed together. The geometry is periodic along $\ez$ with periodicity length $\ell = p/\Ns$, where $p$ is the full transposition length for a strand. The cable carries a transport current $I_\text{t}$ and is subject to a time-varying transverse magnetic field $\happ$, perpendicular to direction $\ez$. 

The CATI method consists in solving separately for (i) the axial currents, flowing in the strands along $\ez$, and (ii) the transverse, or coupling, currents, flowing from one strand to another across contact resistances between them. The periodicity of the geometry is then exploited to couple both models via circuit equations.

\subsection{Axial currents model}

The axial currents model is defined on a 2D cross-section of the cable, such as the one represented in Fig.~\ref{circuitModel}(b). The problem is solved in a computational domain $\O$, decomposed in a conducting domain $\Oc$, containing the strands $\Os$ surrounded by a thin coating domain $\Om$ (introduced for easier definition of currents, as explained later), and the complementary domain $\Occ$, containing the surrounding air or cryogenic liquid.

The model is described by Maxwell's equations in the magnetodynamic regime:
\begin{equation}\label{MQSequations}
\left\{\begin{aligned}
\div\b &= 0,\\
\curl\h &= \j,\\
\curl\e &= -\dt \b,
\end{aligned}\right. \quad \text{with} \quad  \left\{\begin{aligned}
\b &= \mu\, \h,\\
\e &= \rho\, \j,
\end{aligned}\right.
\end{equation}
with $\b$, $\h$, $\j$ $\e$, $\mu$, and $\rho$, the magnetic flux density (T), the magnetic field (A/m), the current density (A/m$^2$), the electric field (V/m), the permeability (H/m), and the resistivity ($\O\,$m), respectively. We assume that the current density in the strands is an axial vector field (along $\ez$), neglecting the tilt angle due to the finite transposition length. The magnetic field is defined as a transverse vector field (perpendicular to $\ez$).

The fields are assumed constant over a length $\ell$ along $\ez$. The fact that strands are transposed will be handled by the circuit coupling equations.

The strands are superconducting but, for simplicity, in this work, we assume a linear Ohm's law for the axial current density, with a small resistivity $\rhos$. This simplification allows to focus on the interstrand coupling current dynamics.

The coating domain $\Om$ helps for the definition of currents with cohomology basis functions. Its resistivity $\rho_\text{m}$ is chosen sufficiently high compared to $\rhos$ in order to have a negligible impact on the solution.

The set of equations~\eqref{MQSequations} is solved with the FE method and the \hpfOnly~\cite{bossavit1998computational}. With the notation $\volInt{\vec f}{\vec g}{\O}$ for the integral over $\O$ of the dot product of any two vector fields $\vec f$ and $\vec g$, the \hpf reads: from an initial solution at $t=0$, find $\h\in\hsp(\O)$ such that, for $t>0$ and $\forall \h' \in \hspz(\O)$, we have~\cite{dular2000dual, dular2024coupled}
\begin{align}\label{eq_oop}
\volInt{\ell\, \dt(\mu\, \h)}{\h'}{\O} + \volInt{\ell\, \rho\, \curl \h}{\curl \h'}{\Oc}\qquad~\notag\\
 = V_\text{t} \mathcal{I}_{\text{t}}(\h') + \sum_{i= 1}^{\Ns} V_i \mathcal{I}_i(\h').
\end{align}
The function space $\hsp(\O)$ is the subspace of $H(\curlOnly;\O)$ containing transverse vector fields that are curl-free in $\Occ$ and fulfill appropriate boundary and global conditions, as defined in~\cite{dular2024coupled}. The space $\hspz(\O)$ is the same space but with homogeneous boundary and global conditions.

The operator $\mathcal{I}_i(\h)$ gives the circulation of $\h$ around strand $i\in \{1,\dots,\Ns\}$, which is the net current $I_i$ flowing in that strand. The associated voltage difference accumulated after length $\ell$ is denoted by $V_i$. The operator $\mathcal{I}_{\text{t}}(\h)$ gives the circulation of $\h$ around the whole cable, i.e., the transport current $I_{\text{t}}$, and $V_{\text{t}}$ denotes the associated voltage difference after length $\ell$. Currents are handled with cohomology basis functions, exactly as was done for the superconducting strand in~\cite{dular2024coupled} and following the procedure described in~\cite{pellikka2013homology}.



\subsection{Transverse currents model}

Contrary to the CATI method applied on a strand for which the transverse currents are described by a FE model in the conducting matrix~\cite{dular2024coupled}, the transverse currents in this Rutherford cable model are defined by lumped resistors associated with the contact surfaces between adjacent and crossing strands. There is therefore no 2D FE model associated with the transverse currents.

Contacts between strands are decomposed in two categories: crossing and adjacent~\cite{verweij1997electrodynamics}. Crossing connections over one periodicity length $\ell$ are represented by the triangles and the rhombuses in the schematic of Fig.~\ref{circuitModel}(a) and also in Fig.~\ref{cable3D}(c). Lumped resistance for currents flowing across the triangles is double that for rhombuses, as it is associated with a surface that is twice smaller. Along length $\ell$, each strand has crossing contact with three other strands, except for strands at the sides of the cable, which only cross two other strands.

Adjacent connections are indicated by the blue curves in Fig.~\ref{circuitModel}(a) and the blue rectangles in Fig.~\ref{cable3D}(c). Each strand has adjacent contact with two other strands.

The connections can be described by lumped contact resistances $\Rc$ and $\Ra$ ($\Omega$), for crossing and adjacent, respectively. In this work, we assume uniform contact resistance values, but the method generally applies for non-uniform values or nonlinear functions as well. Lack of contact between some strands can also be modelled, if needed.


\subsection{Circuit-coupling equations}

The axial currents $I_i$ flowing in the strands and their associated voltages $V_i$ can be coupled with the transverse currents flowing in the lumped resistors and their associated voltages via an electrical circuit as represented in Fig.~\ref{circuitModel}(c). 

The thick black connections in the circuit encode that after periodicity length $\ell$, each strand has taken the position of its adjacent strand. Over this length $\ell$ along $\ez$, the axial currents $I_i$ may increase or decrease as a result of coupling currents through contact resistors.

Interstrand coupling currents are driven by the voltages $V_i$. These voltages are generated by (i) electromotive forces in conducting loops made of pairs of strands connected via transverse currents, and (ii) resistance to axial current flow. Both effects are accounted for by the \hpf in Eq.~\eqref{eq_oop}, as explained in~\cite{dular2000dual, dular2019finite}.

\section{Simulation Results}\label{sec_results}

In this section, we verify the CATI method on a Rutherford cable by comparing its results with those of a reference 3D FE model. As mentioned earlier, we consider a linear problem for simplicity and in order to focus on the interstrand coupling current dynamics.

\begin{figure}[h!]
\begin{center}
\includegraphics[width=0.85\linewidth]{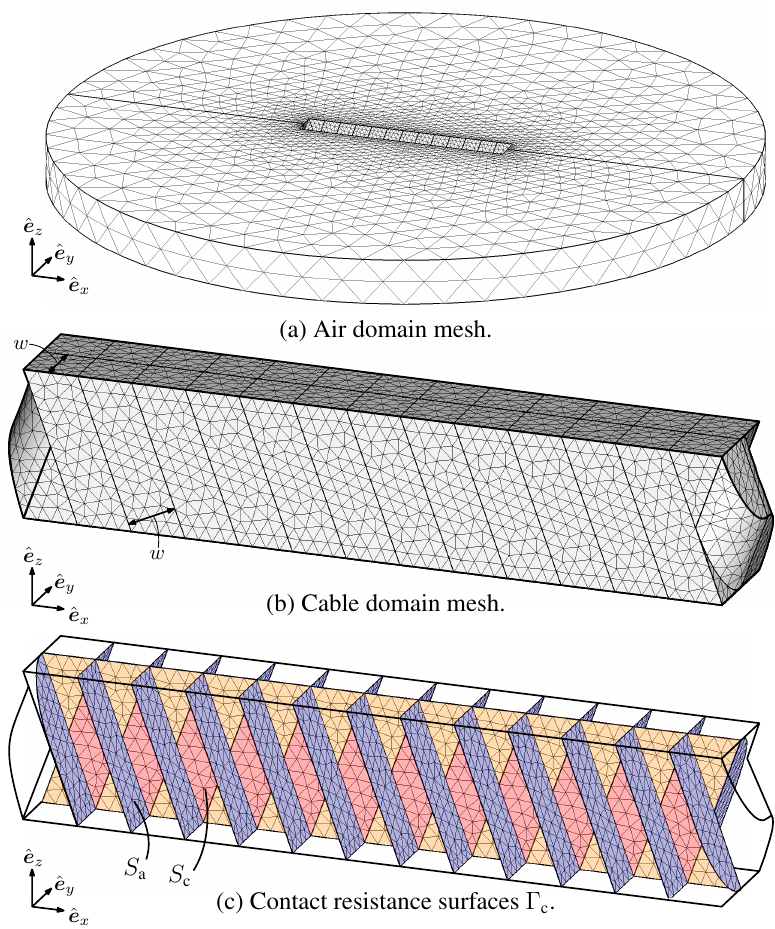}
\caption{FE mesh for the 3D reference model with $\Ns= 26$ strands (the mesh used for results of Section~\ref{sec_results} is much finer). One periodicity length $\ell = p/\Ns$ is modelled with a periodic mesh and periodic boundary conditions on the top and bottom surfaces.} 
\label{cable3D}
\end{center}
\end{figure}

We consider a cable with $\Ns=26$ strands with transposition length $p=0.1$~m, as represented in Fig.~\ref{cable3D}. Strands have a square cross-section of side $w=1.3$~mm. This simple shape is chosen in order to simplify the construction of the 3D geometry, but is not a restriction for the CATI method. The model parameters are given in Table~\ref{cable_parameters}. In the whole domain, we define $\mu = \mu_0 = 4\pi \times 10^{-7}$ H/m.

\begin{table}[!h]\small
\centering
\caption{Parameter values.}
\begin{tabular}{l c r l}
\hline
Number of strands & $\Ns$ & $26$ & -\\
Transposition length & $p$ & $0.1$ &m\\
Strand side length & $w$ & $1.3$ & mm\\ 
Crossing contact resistance & $\Rc$ & $20$ & $\upmu\Omega$\\
Adjacent contact resistance & $\Ra$ & $10$ & $\upmu\Omega$\\
Strand resistivity & $\rho_\text{s}$ & $0.001$ & n$\Omega\,$m\\
Coating matrix resistivity & $\rho_\text{m}$ & $100$ & n$\Omega\,$m\\
\hline
\end{tabular}
\label{cable_parameters}
\end{table}

\subsection{Reference solution}

The reference solution is obtained with a 3D FE model of the cable over one periodicity length $\ell = p/\Ns$. The mesh on the top and bottom surfaces is identical so that periodic boundary conditions can be strongly enforced by constraints on the degrees of freedom of the problem directly. The model is solved with an \hpfOnly~\cite{bossavit1998computational}, with contact resistance modelled via a contact resistivity $r$ (in $\Omega\,$m$^2$) defined on the contact surfaces $\Gamma_\text{c}$ represented in Fig.~\ref{cable3D}(c).

The weak form reads: from an initial solution at $t=0$, find $\h\in\mathcal{H}_{\text{3D}}(\O)$ such that, for $t>0$ and $\forall \h' \in \mathcal{H}_{\text{3D,}0}(\O)$,
\begin{align}\label{eq_hp_contact}
\volInt{\dt(\mu\, \h)}{\h'}{\O} + \volInt{\rho\, \curl \h}{\curl \h'}{\Oc}&\notag\\
+\volInt{(r\, \curl \h)\cdot \n}{(\curl \h')\cdot \n}{\Gamma_\text{c}} &= V_\text{t} \mathcal{I}_{\text{t}}(\h'),
\end{align}
with $\mathcal{H}_{\text{3D}}(\O)$ and $\mathcal{H}_{\text{3D,}0}(\O)$ the usual function spaces for the \hpf in 3D~\cite{dular2019finite}, and $\n$ the unit vector normal to the contact surfaces. 

The local contact resistivity is equal to $r_\text{c} = \Rc S_\text{c}$ on crossing contact surfaces, and $r_\text{a} = \Ra S_\text{a}$ on adjacent contact surfaces, with $S_\text{c}$ and $S_\text{a}$ the surface areas of the red rhombuses and the blue rectangles represented in Fig.~\ref{cable3D}(c), respectively, and with $\Rc$ and $\Ra$ the resistance values used for the CATI model. With the values of Table~\ref{cable_parameters}, this leads to a uniform value $r_\text{c} = r_\text{a} = 5.35\times 10^{-11}$ $\Omega\,$m$^2$. We neglect the fact that the adjacent contact surfaces at the edges of the cable are not exactly rectangles.

\subsection{Result comparison}

The CATI model is defined on a cross-section of the cable which coincides with the top and bottom surfaces of the 3D model. As the problem is linear, it is solved in the frequency domain using complex phasors for the fields, as was done in~\cite{dular2024coupled}. For a fair comparison of the results, the 3D mesh is generated with elements of similar size as those of the 2D mesh for the CATI model. This leads to $1.1\times 10^6$ degrees of freedom (DOF) for the 3D model, compared to only $64\times 10^3$ DOF for the CATI model. Simulations with the CATI method are therefore substantially faster. The computational time for one frequency is around $3$ seconds for the CATI method, compared to more than $4$ minutes in 3D on the same machine.

\begin{figure}[h!]
\begin{center}
\includegraphics[width=0.95\linewidth]{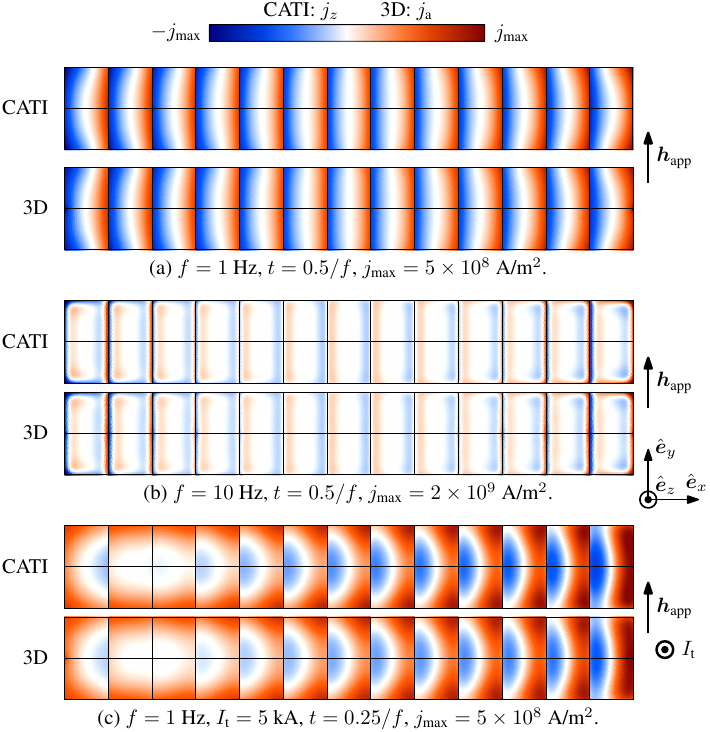}
\caption{Comparison of the current density distribution in three situations. For the CATI method, the axial current density $j_z$ is represented. For the 3D reference solution, the component $j_\text{a}$ of $\j$ along the strand direction is represented. The bottom surface of the model is considered. Bounding values $-j_\text{max}$ and $j_\text{max}$ of the color scale are different in each of the three situations and are indicated in the caption. (a-b) Transverse field along $\ey$ and no transport current, only the frequency differs. (c) Transverse field along $\ey$ and in-phase transport current.}
\label{j_distribution}
\end{center}
\end{figure}

We compare current density distributions obtained with the CATI method with the ones obtained with the 3D reference model. A transverse magnetic field $\happ(t) = \hmax \sin(2\pi f t)\ey$ of amplitude $\mu_0\hmax = 0.1$~T and variable frequency $f$ is applied in the $\ey$ direction. In Fig.~\ref{j_distribution}(a-b), the transport current is fixed to zero, whereas in Fig.~\ref{j_distribution}(c), it is equal to $I_\text{t}(t) = I_\text{t,max} \sin(2\pi f t)$, with $I_\text{t,max} = 5$~kA. The local current density distributions match very well with each other, both with and without transport current.

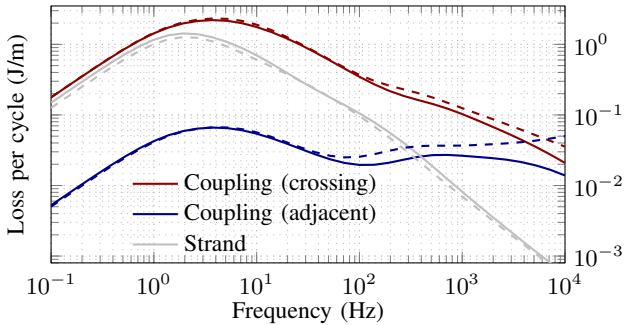
\begin{figure}[h!]
\centering
\tikzsetnextfilename{loss_comparison}
\begin{tikzpicture}[trim axis left, trim axis right][font=\small]
\pgfplotsset{set layers}
 	\begin{loglogaxis}[
	tick label style={/pgf/number format/fixed},
    width=0.95\linewidth,
    height=5cm,
    grid = both,
    grid style = dotted,
    ymin=0.0008, 
    ymax=3.5,
    xmin=0.001, 
    xmax=100,
    xtick={1e-3, 1e-2, 1e-1, 1, 10, 100},
    xticklabels={$10^{-1}$, $10^{0}$, $10^{1}$, $10^{2}$, $10^{3}$, $10^{4}$ },
	xlabel={Frequency (Hz)},
    ylabel={Loss per cycle (J/m)},
    ylabel style={yshift=-2.5em},
    xlabel style={yshift=0.4em},
    xticklabel style={yshift=-0.1em},
    yticklabel style={xshift=0em},
    yticklabel pos=right,
    legend style={at={(0.4, -0.01)}, cells={anchor=west}, anchor=south, draw=none,fill opacity=0, text opacity = 1}]
        \addplot[darkred, thick, solid] 
    table[x=f,y=crossing_2D]{data/loss_comparison.txt};
        \addplot[darkblue, thick, solid]
    table[x=f,y=adjacent_2D]{data/loss_comparison.txt};
        \addplot[lightgray, thick, solid] 
    table[x=f,y=strand_2D]{data/loss_comparison.txt};
        \addplot[darkred, thick, dashed] 
    table[x=f,y=crossing_3D]{data/loss_comparison.txt};
        \addplot[darkblue, thick, dashed]
    table[x=f,y=adjacent_3D]{data/loss_comparison.txt};
        \addplot[lightgray, thick, dashed] 
    table[x=f,y=strand_3D]{data/loss_comparison.txt};
    \legend{Coupling (crossing), Coupling (adjacent), Strand}
    \end{loglogaxis}
\end{tikzpicture}%
\vspace{-0.2cm}
\caption{Comparison of the loss per cycle and per unit length between the CATI method (solid curves) and the 3D reference model (dashed curves) as a function of frequency for a transverse field excitation of amplitude $\mu_0\hmax = 0.1$~T along $\ey$. The total coupling loss is the sum of the crossing coupling current loss (red) and the adjacent coupling current loss (blue). The strand loss is the Joule loss computed in the strands, due to the resistivity $\rho_\text{s}$.}
\label{loss_comparison}
\end{figure}

Figure~\ref{loss_comparison} gives the loss per cycle and per unit length along the cable. The maximum coupling loss values differ by less than $5$\%, and are observed at almost identical frequencies $f_\text{c}\approx 3.5$~Hz, indicating that the coupling current dynamics is well reproduced with the CATI method. The associated time constant $\tau_\text{c} = 1/(2\pi f_\text{c}) = 45$~ms also matches calculations from network models, e.g., in~\cite{verweij1997electrodynamics} it is given by
\begin{align}
\tau_\text{c} = C\,\frac{p\, (\Ns^2 - \Ns)}{\Rc} = 47\ \text{ms}, 
\end{align}
with $C = 1.65\times 10^{-8}$ $\Omega\,$s\,m$^{-1}$.

\begin{figure}[h!]
\centering
\subfloat{%
\tikzsetnextfilename{coupling_currents_imag}
\begin{tikzpicture}[trim axis left, trim axis right][font=\small]
\pgfplotsset{set layers}
 	\begin{axis}[
	tick label style={/pgf/number format/fixed},
    width=0.95\linewidth,
    height=4.3cm,
    grid = both,
    grid style = dotted,
    ymin=-30, 
    ymax=50,
    xmin=-1, 
    xmax=1,
    xticklabels={},
    ylabel={Current (A)},
    ylabel style={yshift=-2.5em},
    ylabel style={xshift=0em},
    xlabel style={yshift=0.4em},
    xticklabel style={yshift=-0.4em},
    yticklabel style={xshift=0em},
    yticklabel pos=right,
    legend columns=3,
    transpose legend,
    legend style={at={(0.5, 0.99)}, cells={anchor=west}, anchor=north, draw=none,fill opacity=0, text opacity = 1}
    ]
    \addplot[vir_1, thick, solid, mark=*, mark options={vir_1, scale=0.5, style={solid}}] 
    table[x=position,y=current_2D]{data/coupling_crossing_currents_imag_f0.001.txt};

        \addplot[vir_2, thick, solid, mark=*, mark options={vir_2, scale=0.5, style={solid}}]
    table[x=position,y=current_2D]{data/coupling_crossing_currents_imag_f0.01.txt};
        \addplot[vir_3, thick, solid, mark=*, mark options={vir_3, scale=0.5, style={solid}}]
    table[x=position,y=current_2D]{data/coupling_crossing_currents_imag_f0.1.txt};
        \addplot[vir_4, thick, solid, mark=*, mark options={vir_4, scale=0.5, style={solid}}]
    table[x=position,y=current_2D]{data/coupling_crossing_currents_imag_f1.0.txt};
        \addplot[vir_5, thick, solid, mark=*, mark options={vir_5, scale=0.5, style={solid}}]
    table[x=position,y=current_2D]{data/coupling_crossing_currents_imag_f10.0.txt};
        \addplot[vir_6, thick, solid, mark=*, mark options={vir_6, scale=0.5, style={solid}}]
    table[x=position,y=current_2D]{data/coupling_crossing_currents_imag_f100.0.txt};
    \addplot[vir_1, thick, densely dashed, mark=*, mark options={vir_1, scale=0.5, style={solid}}]
    table[x=position,y=current_3D]{data/coupling_crossing_currents_imag_f0.001.txt};
     \addplot[vir_2, thick, densely dashed, mark=*, mark options={vir_2, scale=0.5, style={solid}}]
    table[x=position,y=current_3D]{data/coupling_crossing_currents_imag_f0.01.txt};
     \addplot[vir_3, thick, densely dashed, mark=*, mark options={vir_3, scale=0.5, style={solid}}]
    table[x=position,y=current_3D]{data/coupling_crossing_currents_imag_f0.1.txt};  
    \addplot[vir_4, thick, densely dashed, mark=*, mark options={vir_4, scale=0.5, style={solid}}]
    table[x=position,y=current_3D]{data/coupling_crossing_currents_imag_f1.0.txt}; 
    \addplot[vir_5, thick, densely dashed, mark=*, mark options={vir_5, scale=0.5, style={solid}}]
    table[x=position,y=current_3D]{data/coupling_crossing_currents_imag_f10.0.txt};
    \addplot[vir_6, thick, densely dashed, mark=*, mark options={vir_6, scale=0.5, style={solid}}]
    table[x=position,y=current_3D]{data/coupling_crossing_currents_imag_f100.0.txt};
    \node[anchor=center] at (axis cs: 0, -20) {Crossing coupling currents};
    \legend{$0.1$ Hz, $1$ Hz, $10$ Hz, $100$ Hz, $1\, 000$ Hz, $10\, 000$ Hz}
    \end{axis}
\end{tikzpicture}%
}
\vspace{-0.5cm}
\hfill
\subfloat{%
\tikzsetnextfilename{adjacent_currents_imag}
\begin{tikzpicture}[trim axis left, trim axis right][font=\small]
\pgfplotsset{set layers}
 	\begin{axis}[
	tick label style={/pgf/number format/fixed},
    width=0.95\linewidth,
    height=4cm,
    grid = both,
    grid style = dotted,
    ymin=-30, 
    ymax=30,
    xmin=-1, 
    xmax=1,
    xticklabels={},
    ylabel={Current (A)},
    ylabel style={yshift=-2.5em},
    ylabel style={xshift=0em},
    xlabel style={yshift=0.4em},
    xticklabel style={yshift=-0.4em},
    yticklabel style={xshift=0em},
    yticklabel pos=right,
    legend style={at={(0.33, 1.0)}, cells={anchor=west}, anchor=north, draw=none,fill opacity=0, text opacity = 1}
    ]
    \addplot[vir_1, thick, solid, mark=*, mark options={vir_1, scale=0.5, style={solid}}] 
    table[x=position,y=current_2D]{data/coupling_adjacent_currents_imag_f0.001.txt};
        \addplot[vir_2, thick, solid, mark=*, mark options={vir_2, scale=0.5, style={solid}}]
    table[x=position,y=current_2D]{data/coupling_adjacent_currents_imag_f0.01.txt};
        \addplot[vir_3, thick, solid, mark=*, mark options={vir_3, scale=0.5, style={solid}}]
    table[x=position,y=current_2D]{data/coupling_adjacent_currents_imag_f0.1.txt};
        \addplot[vir_4, thick, solid, mark=*, mark options={vir_4, scale=0.5, style={solid}}]
    table[x=position,y=current_2D]{data/coupling_adjacent_currents_imag_f1.0.txt};
        \addplot[vir_5, thick, solid, mark=*, mark options={vir_5, scale=0.5, style={solid}}]
    table[x=position,y=current_2D]{data/coupling_adjacent_currents_imag_f10.0.txt};
        \addplot[vir_6, thick, solid, mark=*, mark options={vir_6, scale=0.5, style={solid}}]
    table[x=position,y=current_2D]{data/coupling_adjacent_currents_imag_f100.0.txt};
    \addplot[vir_1, thick, densely dashed, mark=*, mark options={vir_1, scale=0.5, style={solid}}]
    table[x=position,y=current_3D]{data/coupling_adjacent_currents_imag_f0.001.txt};
    \addplot[vir_2, thick, densely dashed, mark=*, mark options={vir_2, scale=0.5, style={solid}}]
    table[x=position,y=current_3D]{data/coupling_adjacent_currents_imag_f0.01.txt};
    \addplot[vir_3, thick, densely dashed, mark=*, mark options={vir_3, scale=0.5, style={solid}}]
    table[x=position,y=current_3D]{data/coupling_adjacent_currents_imag_f0.1.txt};
    \addplot[vir_4, thick, densely dashed, mark=*, mark options={vir_4, scale=0.5, style={solid}}]
    table[x=position,y=current_3D]{data/coupling_adjacent_currents_imag_f1.0.txt};
    \addplot[vir_5, thick, densely dashed, mark=*, mark options={vir_5, scale=0.5, style={solid}}]
    table[x=position,y=current_3D]{data/coupling_adjacent_currents_imag_f10.0.txt};
    \addplot[vir_6, thick, densely dashed, mark=*, mark options={vir_6, scale=0.5, style={solid}}]
    table[x=position,y=current_3D]{data/coupling_adjacent_currents_imag_f100.0.txt};
        \node[anchor=center] at (axis cs: -0.1, -22) {Adjacent coupling currents};
    \legend{}
    \end{axis}
\end{tikzpicture}%
}
\vspace{-0.7cm}
\hfill
\subfloat{%
\centering
\tikzsetnextfilename{transport_currents_imag}
\begin{tikzpicture}[trim axis left, trim axis right][font=\small]
\pgfplotsset{set layers}
 	\begin{axis}[
	tick label style={/pgf/number format/fixed},
    width=0.95\linewidth,
    height=4cm,
    grid = both,
    grid style = dotted,
    ymin=-100, 
    ymax=100,
    xmin=-1, 
    xmax=1,
    xtick={-1, -0.5, 0, 0.5, 1},
    xticklabels={$-9$, $-4.5$, $0$, $4.5$, $9$},
	xlabel={$x$ (mm)},
    ylabel={Current (A)},
    ylabel style={yshift=-2.5em},
    xlabel style={yshift=0.6em},
    xticklabel style={yshift=0.0em},
    yticklabel style={xshift=0em},
    yticklabel pos=right,
    legend columns=3,
    transpose legend,
    legend style={at={(0.72, -0.05)}, cells={anchor=west}, anchor=south, draw=none,fill opacity=0, text opacity = 1}
    ]
    \addplot[vir_1, thick, solid, mark=*, mark options={vir_1, scale=0.5, style={solid}}] 
    table[x=position,y=current_2D]{data/strand_currents_imag_f0.001.txt};
        \addplot[vir_2, thick, solid, mark=*, mark options={vir_2, scale=0.5, style={solid}}]
    table[x=position,y=current_2D]{data/strand_currents_imag_f0.01.txt};
        \addplot[vir_3, thick, solid, mark=*, mark options={vir_3, scale=0.5, style={solid}}]
    table[x=position,y=current_2D]{data/strand_currents_imag_f0.1.txt};
        \addplot[vir_4, thick, solid, mark=*, mark options={vir_4, scale=0.5, style={solid}}]
    table[x=position,y=current_2D]{data/strand_currents_imag_f1.0.txt};
        \addplot[vir_5, thick, solid, mark=*, mark options={vir_5, scale=0.5, style={solid}}]
    table[x=position,y=current_2D]{data/strand_currents_imag_f10.0.txt};
        \addplot[vir_6, thick, solid, mark=*, mark options={vir_6, scale=0.5, style={solid}}]
    table[x=position,y=current_2D]{data/strand_currents_imag_f100.0.txt};
    \addplot[vir_1, thick, densely dashed, mark=*, mark options={vir_1, scale=0.5, style={solid}}]
    table[x=position,y=current_3D]{data/strand_currents_imag_f0.001.txt};    
        \addplot[vir_2, thick, densely dashed, mark=*, mark options={vir_2, scale=0.5, style={solid}}]
    table[x=position,y=current_3D]{data/strand_currents_imag_f0.01.txt};
        \addplot[vir_3, thick, densely dashed, mark=*, mark options={vir_3, scale=0.5, style={solid}}]
    table[x=position,y=current_3D]{data/strand_currents_imag_f0.1.txt};
        \addplot[vir_4, thick, densely dashed, mark=*, mark options={vir_4, scale=0.5, style={solid}}]
    table[x=position,y=current_3D]{data/strand_currents_imag_f1.0.txt};
        \addplot[vir_5, thick, densely dashed, mark=*, mark options={vir_5, scale=0.5, style={solid}}]
    table[x=position,y=current_3D]{data/strand_currents_imag_f10.0.txt};
    \addplot[vir_6, thick, densely dashed, mark=*, mark options={vir_6, scale=0.5, style={solid}}]
    table[x=position,y=current_3D]{data/strand_currents_imag_f100.0.txt};
            \node[anchor=center] at (axis cs: 0, -75) {Strand currents $I_i$};
    \legend{}
    \end{axis}
\end{tikzpicture}%
}
\vspace{-0.2cm}
\caption{Comparison of currents between the CATI method (solid lines) and 3D reference model (dashed lines), for a transverse field excitation of amplitude $\mu_0\hmax = 0.1$~T along $\ey$ and different frequencies (same legend for the three subfigures). Solutions at $t=0.5/f$. Shown crossing currents are those across the red resistors in Fig.~\ref{circuitModel} or the red rhombuses of Fig.~\ref{cable3D}. As the transport current is zero, the solution is identical on the top and bottom layers (up to a sign difference depending on the conventions). Hence, solutions of only one layer are represented.}
\label{coupling_currents_imag}
\end{figure}
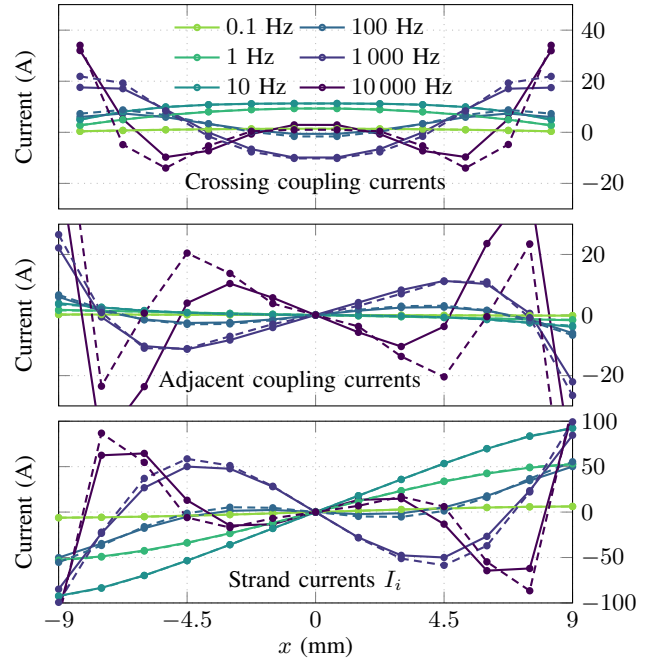

Figure~\ref{coupling_currents_imag} shows the coupling and strand currents at different frequencies for both models. For $f\lesssim 1$~kHz, the results of both models agree very well with each other. At low frequencies, $f\lesssim f_\text{c}$, strands are mostly not coupled. As the frequency increases, coupling increases, which results in higher strand currents $I_i$. For frequencies $f \gtrsim 1$~kHz, screening current effects become visible. The agreement between the models deteriorates at these frequencies, especially for the adjacent coupling currents. The relevance of using a linear model in that regime is however questionable, as the skin depth becomes very small compared to the strands (at $f=10$~kHz, the skin depth is $4$~$\upmu$m), such that edge effects in the strands on the sides of the cables (see Fig.~\ref{cable3D}) become dominant. These geometrical effects are not considered in the CATI model.

\FloatBarrier
\section{Conclusion}

In this paper, we applied the CATI method approach to Rutherford cables and verified it by comparison with a reference 3D FE model in a linear setting. The CATI method was shown to correctly reproduce the interstrand coupling currents and the associated loss of the 3D model, with a substantially reduced computational cost. It allows for detailed and fast analyses of arbitrary cable geometries.

The next step consists in applying the method on a nonlinear problem that accounts for the complex dynamics of currents in the strands, with either detailed strand models in each of them, or homogenized material properties obtained from preliminary simulations on reference strands. Such a model would include all loss contributions down to the filament level.

\clearpage
\newpage

\bibliographystyle{ieeetr}
\bibliography{../../paperReferences}

\begin{thebibliography}{10}

\bibitem{dangelo2021quasi}
L.~A. D'Angelo and H.~De~Gersem, ``Quasi-{3D} magnetic field simulation of
  superconducting devices with translational symmetry,'' {\em IET Science,
  Measurement \& Technology}, vol.~15, no.~3, pp.~319--327, 2021.

\bibitem{riva2023h}
N.~Riva, A.~Halbach, M.~Lyly, C.~Messe, J.~Ruuskanen, and V.~Lahtinen,
  ``{H}-phi formulation in {S}parselizard combined with domain decomposition
  methods for modeling superconducting tapes, stacks, and twisted wires,'' {\em
  IEEE Transactions on Applied Superconductivity}, vol.~33, no.~5, pp.~1--5,
  2023.

\bibitem{wilson1983superconducting}
M.~N. Wilson, ``Superconducting magnets,'' {\em Clarendon Press, United
  Kingdom}, 1983.

\bibitem{mulder2023external}
T.~Mulder, B.~Bordini, E.~Ravaioli, E.~Schnaubelt, M.~Wozniak, and A.~Verweij,
  ``External coil coupled loss induced quench ({E-CLIQ}) system for the
  protection of {LTS} magnets,'' {\em IEEE Transactions on Applied
  Superconductivity}, vol.~33, no.~5, pp.~1--5, 2023.

\bibitem{ravaioli2023optimizing}
E.~Ravaioli, T.~Mulder, A.~Verweij, and M.~Wozniak, ``Optimizing secondary
  {CLIQ} for protecting high-field accelerator magnets,'' {\em IEEE
  Transactions on Applied Superconductivity}, vol.~34, no.~5, pp.~1--5, 2024.

\bibitem{morgan1970theoretical}
G.~Morgan, ``Theoretical behavior of twisted multicore superconducting wire in
  a time-varying uniform magnetic field,'' {\em Journal of Applied Physics},
  vol.~41, no.~9, pp.~3673--3679, 1970.

\bibitem{ries1981coupling}
G.~Ries and S.~Tak{\'a}cs, ``Coupling losses in finite length of
  superconducting cables and in long cables partially in magnetic field,'' {\em
  IEEE Transactions on Magnetics}, vol.~17, no.~5, pp.~2281--2284, 1981.

\bibitem{verweij1997electrodynamics}
A.~P. Verweij, ``Electrodynamics of superconducting cables in accelerator
  magnets.,'' {\em PhD thesis, Twente University, Enschede}, 1997.

\bibitem{krempasky1998influence}
L.~Krempasky and C.~Schmidt, ``Influence of supercurrents on the stability of
  superconducting magnets,'' {\em Physica C: Superconductivity}, vol.~310,
  no.~1-4, pp.~327--334, 1998.

\bibitem{bottura2018calculation}
L.~Bottura, M.~Breschi, and A.~Musso, ``Calculation of interstrand coupling
  losses in superconducting rutherford cables with a continuum model,'' {\em
  Cryogenics}, vol.~96, pp.~44--52, 2018.

\bibitem{dular2024coupled}
J.~Dular, F.~Magnus, E.~Schnaubelt, A.~Verweij, and M.~Wozniak, ``Coupled axial
  and transverse currents method for finite element modelling of periodic
  superconductors,'' {\em Superconductor Science and Technology}, vol.~37,
  no.~9, pp.~1--18, 2024.

\bibitem{gmsh}
C.~Geuzaine and J.-F. Remacle, ``{G}msh: A 3{D} finite element mesh generator
  with built-in pre-and post-processing facilities,'' {\em International
  journal for numerical methods in engineering}, vol.~79, no.~11,
  pp.~1309--1331, 2009.

\bibitem{getdp}
P.~Dular, C.~Geuzaine, F.~Henrotte, and W.~Legros, ``A general environment for
  the treatment of discrete problems and its application to the finite element
  method,'' {\em IEEE Transactions on Magnetics}, vol.~34, no.~5,
  pp.~3395--3398, 1998.

\bibitem{vitrano2023open}
A.~Vitrano, M.~Wozniak, E.~Schnaubelt, T.~Mulder, E.~Ravaioli, and A.~Verweij,
  ``An open-source finite element quench simulation tool for superconducting
  magnets,'' {\em IEEE Transactions on Applied Superconductivity}, vol.~33,
  no.~5, pp.~1--6, 2023.

\bibitem{Bortot2017}
L.~Bortot, B.~Auchmann, I.~C. Garcia, A.~F. Navarro, M.~Maciejewski,
  M.~Mentink, M.~Prioli, E.~Ravaioli, S.~Schoeps, and A.~Verweij, ``{STEAM}: A
  hierarchical cosimulation framework for superconducting accelerator magnet
  circuits,'' {\em IEEE Transactions on Applied Superconductivity}, vol.~28,
  no.~3, pp.~1--6, 2017.

\bibitem{bossavit1998computational}
A.~Bossavit, {\em Computational electromagnetism: variational formulations,
  complementarity, edge elements}.
\newblock Academic Press, 1998.

\bibitem{dular2000dual}
P.~Dular, P.~Kuo-Peng, C.~Geuzaine, N.~Sadowski, and J.~Bastos, ``Dual
  magnetodynamic formulations and their source fields associated with massive
  and stranded inductors,'' {\em IEEE Transactions on Magnetics}, vol.~36,
  no.~4, pp.~1293--1299, 2000.

\bibitem{pellikka2013homology}
M.~Pellikka, S.~Suuriniemi, L.~Kettunen, and C.~Geuzaine, ``Homology and
  cohomology computation in finite element modeling,'' {\em SIAM Journal on
  Scientific Computing}, vol.~35, no.~5, pp.~B1195--B1214, 2013.

\bibitem{dular2019finite}
J.~Dular, C.~Geuzaine, and B.~Vanderheyden, ``Finite-element formulations for
  systems with high-temperature superconductors,'' {\em IEEE Transactions on
  Applied Superconductivity}, vol.~30, no.~3, pp.~1--13, 2019.

\end{thebibliography}

\end{document}